# On-liquid-gallium surface synthesis of ultra-smooth conductive metal-organic framework thin films

Jinxin Liu[1,§], Yunxu Chen[2,§], Xing Huang[2], Yanhan Ren[3], Mike Hambsch[4], David Bodesheim[5], Darius Pohl[6], Xiaodong Li[2], Marielle Deconinck[7], Bowen Zhang[8], Markus Löffler[6], Zhongquan Liao[8], Fengxiang Zhao[9], Arezoo Dianat[5], Gianaurelio Cuniberti[5], Yana Vaynzof[7], Junfeng Gao[3], Jingcheng Hao[9], Stefan C. B. Mannsfeld[4], Xinliang Feng[1,2,*], Renhao Dong[1,9,*]

*[1]Center for Advancing Electronics Dresden (CFAED) & Faculty of Chemistry and Food Chemistry, Technische Universität Dresden, 01062 Dresden, Germany. [2]Max Planck Institute for Microstructure Physics, 06120 Halle (Saale), Germany. [3]Laboratory of Materials Modification by Laser, Ion and Electron Beams, Ministry of Education, Dalian University of Technology, 116024 Dalian, P. R. China. [4]Center for Advancing Electronics Dresden (CFAED) and Faculty of Electrical and Computer Engineering, Technische Universität Dresden, 01062 Dresden, Germany. [5]Institute for Materials Science and Max Bergmann Center of Biomaterials, Technische Universität Dresden, 01062 Dresden, Germany. [6]Dresden Center for Nanoanalysis (DCN), Center for Advancing Electronics Dresden (CFAED), Technische Universität Dresden, 01069 Dresden, Germany. [7] Integrated Centre for Applied Physics and Photonic Materials (IAPP), TUD Dresden University of Technology, Dresden, Germany [8]Fraunhofer Institute for Ceramic Technologies and Systems (IKTS), 01109 Dresden, Germany. [9] Key Laboratory of Colloid and Interface Chemistry of the Ministry of Education, School of Chemistry and Chemical Engineering, Shandong University, Jinan 250100, China.*

*\*Corresponding author: xinliang.feng@tu-dresden.de, renhaodong@sdu.edu.cn*

[§]These authors contributed equally to this work.

**Abstract**

Conductive metal-organic frameworks (MOFs) are emerging electroactive materials for (opto-)electronics. However, it remains a great challenge to achieve reliable MOF-based devices via the existing synthesis methods that are compatible with the complementary metal-oxide-semiconductor technology, as the surface roughness of thus-far synthetic MOF films or pellets is rather high for efficient electrode contact. Here, we develop an on-liquid-gallium surface synthesis (OLGSS) strategy under chemical vapor deposition (CVD) conditions for the controlled growth of two-dimensional conjugated MOF (2D $c$-MOF) thin films with ten-fold improvement of surface flatness (surface roughness can reach as low as ~2 Å) compared with MOF films grown by the traditional methods. Supported by theoretical modeling, we unveil a layer-by-layer CVD growth mode for constructing flattening surfaces, that is triggered by the high adhesion energy between gallium (Ga) and planar aromatic ligands. We further demonstrate the generality of the as-proposed OLGSS strategy by reproducing such a flat surface over nine different 2D $c$-MOF films with variable thicknesses (~2 to 208 nm) and large lateral sizes (over 1 cm$^2$). The resultant ultra-smooth 2D $c$-MOF films enable the formation of high-quality electrical contacts with gold (Au) electrodes, leading to a reduction of contact resistance by over ten orders of magnitude compared to the traditional uneven MOF films. Furthermore, due to the efficient interfacial interaction benifited from the high-quality contacts, the prepared van der Waals heterostructure (vdWH) of OLGSS $c$-MOF and MoS$_2$ exhibits intriguing photoluminescence (PL) enhancement, PL peak shift and large work function modulation. The establishment of the reliable OLGSS method provides the chances to push the development of MOF electronics and the construction of multicomponent MOF-based heterostructure materials.

Electrically conductive metal-organic frameworks (MOFs) or crystalline coordination polymers have recently received considerable attention as attractive electroactive materials[1–4]. Two-dimensional conjugated MOFs (2D *c*-MOFs)[5,6] refer to a class of layer-stacked conductive MOFs linked by square-planar complexes with in-plane $\pi$-extended conjugation and out-of-plane van der Waals interaction. The unique charge transport properties[7,8] and the structural/compositional diversity result in 2D *c*-MOFs with numerous intriguing physical properties such as high intrinsically electrical conductivities (up to $10^3$ S/cm)[9,10], tailorable band gaps from metal to semiconductor[11], high charge mobility (up to $10^2$ cm$^2$/Vs)[12], photosensitivity[13], thermoelectricity[14], superconductivity[15], ferromagnetic ordering[10] and topological state[16], which enable 2D *c*-MOFs for broad (opto-)electronic applications.

In order to integrate 2D *c*-MOFs into reliable (opto-)electronic devices, great efforts have been made to develop synthetic methodologies that yield thin film samples compatible with current complementary metal-oxide-semiconductor (CMOS) technology[17]. Reassembly of the exfoliated nanosheets has been taken as a straightforward approach to fabricate 2D *c*-MOF films[18], but was limited by the poorly controlled film crystallinity and thicknesses[6]. The on-water surface synthesis[19,20] and liquid-interface-assisted synthesis (LIAS) methods enable the growth of 2D *c*-MOF films with variable thicknesses and allow feasible transfer for device integrations[6,21]. However, the residual contamination and the formation of coordinated particles that are dictated by the poor controllability of precursor nucleation lead to rough film surface[22,23] (root-mean-square roughness ($R_q$) ranges from 20 to $10^3$ Å), thus compromising the interfacial charge transport. Recently, solvent-free chemical vapor deposition (CVD) approaches, known for the synthesis of traditional inorganic 2D materials[24] such as graphene and MoS$_2$, have been utilized for the growth of 2D *c*-MOF thin films on silicon dioxide (SiO$_2$) surfaces[25–27], yet still presented high $R_q$ values up to 170 Å. Such high surface roughness results from similar adsorption strength of planar aromatic ligands adopting edge-on (the substitution groups) and face-on (the aromatic planes) orientations on the SiO$_2$ surface[28], thus leading to the disordered arrangement of ligands, which also limits the further formation of smooth 2D *c*-MOF films. Clearly, the further integration of these reported 2D *c*-MOF films for (opto-)electronics still encounters great challenges as the surface roughness is too high for efficient electrode contact[29,30], which urgently requires the development of a new synthetic strategy to realize high surface flatness.

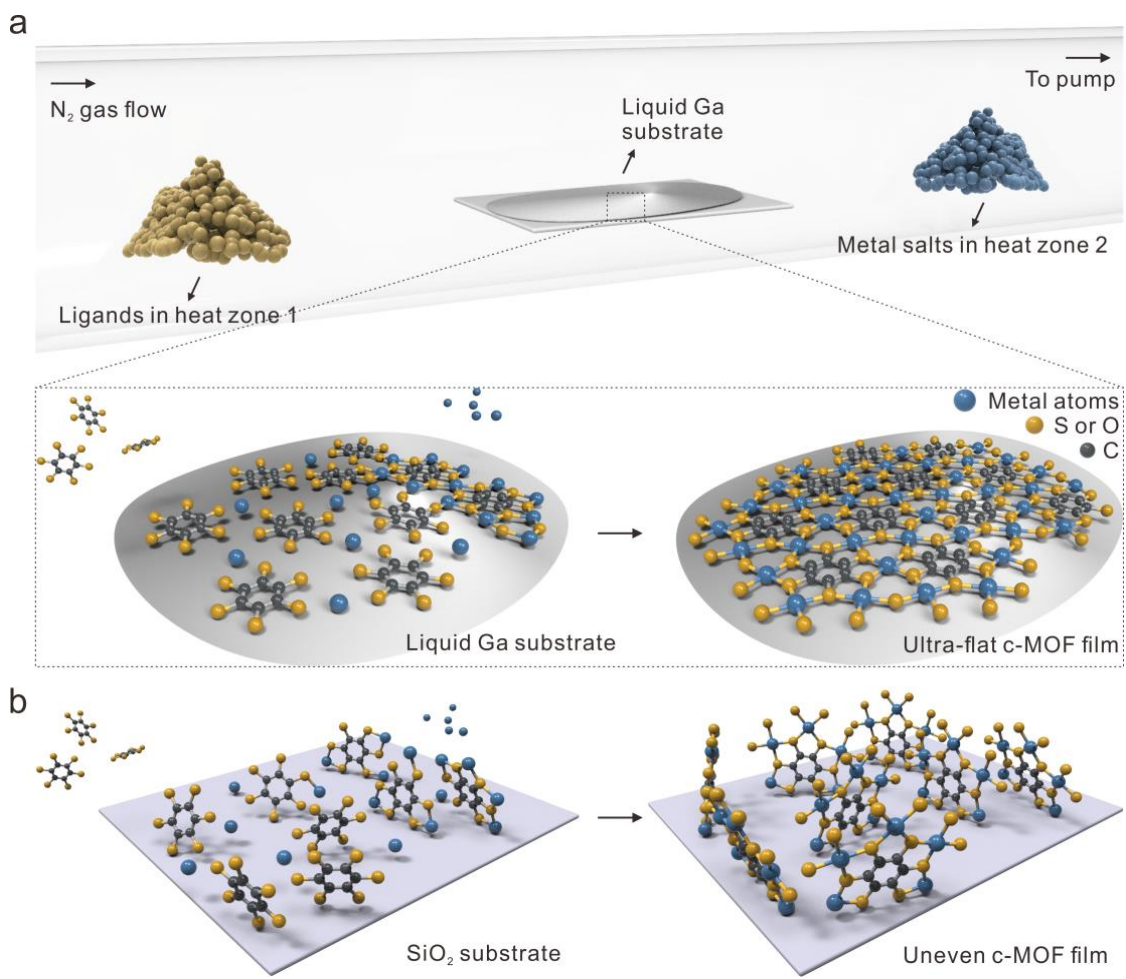

**Fig. 1 | Schematic illustration of the synthesis of 2D *c*-MOF films on different substrates under CVD condition.** (a) A scheme illustrating the typical CVD setup to perform the OLGSS process. The inset at the bottom shows the assembly and the subsequent polymerization of ligands with metal atoms into ultra-smooth 2D *c*-MOFs thin films on liquid Ga substrate. (b) Schematic illustration of the synthesis process of uneven 2D *c*-MOF thin films, where the ligands present disordered arrangement on SiO₂ substrate.

Herein, we develop a general CVD method based on an on-liquid-gallium surface synthesis (OLGSS) strategy (Fig. 1a) for the growth of nine representative ultra-smooth 2D *c*-MOF thin films with the $R_q$ down to ~2 Å, signifying at least ten-fold improvement in surface flatness compared to the reported MOF films (20 to $10^3$ Å)[9,25–27,31–34]. Different from the SiO₂ surface that presents an intense electrostatic effect with hydroxyl groups[28], the liquid gallium (Ga) with atomic-smooth surface[35,36] has demonstrated potential effective interaction with the planar aromatic system, such as graphene[37], thereby holding great promise to realize ultra-smooth 2D *c*-MOF thin films (Fig. 1a and b). As proposed by the theoretical calculations, the surpassing adsorption capacity of the adsorbed aromatic ligands with face-on orientation on the Ga surface ensures the formation of the initial flat 2D $c$-MOF layer. Furthermore, the high adhesion energy between 2D $c$-MOF layers and Ga permits the film growth following the layer-by-layer mode, thereby yielding a uniform thin film with ultra-smooth surface. Owing to the highly improved interface contact, the contact resistance in the integrated devices of OLGSS films can be reduced by more than 10 times compared to that of the samples grown on SiO$_2$ surface. In addition, the van der Waals heterostructures (vdWHs) constructed by few-layer OLGSS Cu-BHT and monolayer MoS$_2$ reveal large modulations of the photoluminescence (PL) properties and work function (WF), which highlights the potential of OLGSS films as building blocks for creating unique vdWHs and uncovering novel interfacial phenomena. Our work develops a reliable OLGSS strategy that affords a feasible access to versatile operable 2D $c$-MOF films with ultra-smooth surfaces as well as their vdWHs and breaks one synthetic bottleneck of conductive MOFs for (opto-)electronic devices.

## Results and discussion

**Synthesis and characterization of ultra-smooth 2D $c$-MOFs.** The OLGSS was performed in a typical tube furnace with two-heating zones (as schemed in Fig. 1a and Supplementary Fig. 1). Before triggering the coordination reaction, the chamber was first purged by N$_2$ and pumped to ~0.5 mbar to facilitate evaporation of precursors. Our general synthesis protocol is shown in Fig. 2a. Various 2D $c$-MOF films were synthesized by employing corresponding benzene-based ligands, including 1,2,3,4,5,6-hexahydroxybenzene (HHB, C$_6$H$_6$O$_6$) and 1,2,3,4,5,6-benzenehexathiol (BHT, C$_6$H$_6$S$_6$), to coordinate with six metal centers (M) such as copper (Cu), nickel (Ni), cobalt (Co), palladium (Pd), vanadium (V) and platinum (Pt). In a typical synthesis procedure of Cu-BHT (Fig. 2b), the BHT ligands and Cu(II) acetylacetonate powders were heated to 130 and 90 °C, respectively, in 10 min, to evaporate the reactants. The precursor vapors flowed to the Ga substrate in such a low-pressure environment and initiated the coordination reaction. After a reaction time of 10 min to 12 hours, the ultra-smooth Cu-BHT films with a thickness ranging from ~ 1.7 nm to ~ 208 nm were formed on the liquid Ga surface. The uniformity and surface flatness of the synthesized OLGSS Cu-BHT (named as O-Cu-BHT) films were initially examined by scanning electron microscopy (SEM). The typical low-magnification and high-magnification SEM images of the sample intuitively indicate that the O-Cu-BHT thin film is highly homogeneous in a large area without observation of any unevenness (Supplementary Fig. 2). For further characterization by atomic force microscopy (AFM), the O-Cu-BHT thin films were transferred onto $SiO_2$/Si substrates assisted by commercial thermal release tapes (Supplementary Fig. 3 and 4)[38]. After thoroughly rinsing off residual Ga with diluted hydrochloric acid (pH = 1~2), the subsequent thermal releasing process affords large-area uniform O-Cu-BHT thin films. The AFM topographic image shows that the surface roughness $R_q$ of the O-Cu-BHT film over an area of $25 \times 25 \ \mu m^2$ was measured as 2.51 Å (Fig. 2c), which is at least 16 times smoother than those of the Cu-BHT films by traditional synthesis (42 ~ 170 Å)[26].

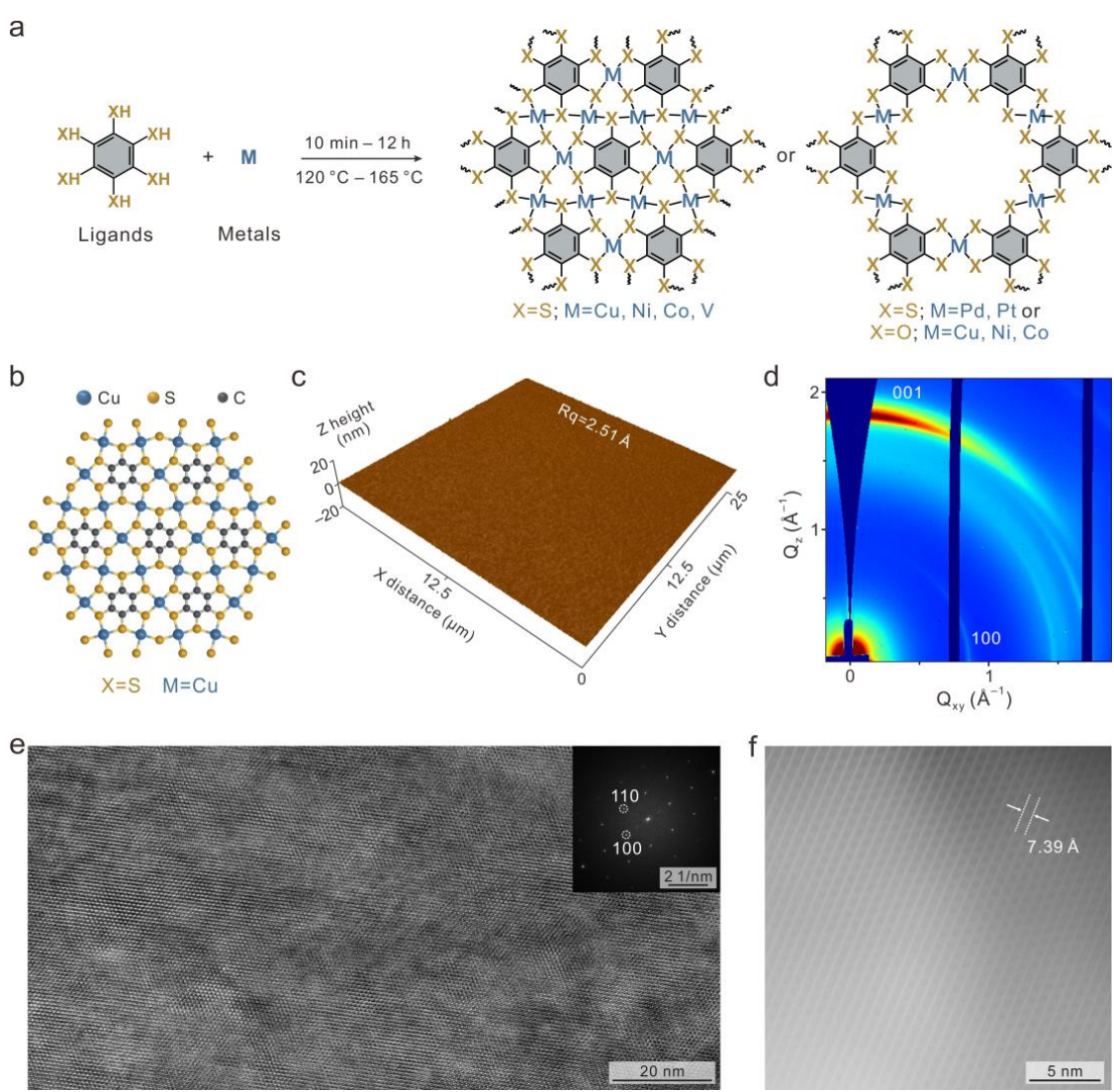

**Fig. 2 | OLGSS 2D *c*-MOF thin films with an ultra-smooth surface.** (a) A reaction scheme illustrating the synthesis of OLGSS 2D *c*-MOFs via coordination between different organic ligands and metals, where the organic ligands include HHB and BHT

and the metals include Cu, Ni, Co, Pd, V and Pt. (b) Atomic model of the Cu-BHT. (c) 3D representations of the AFM topographic data for the O-Cu-BHT thin film by OLGSS. (d) The 2D GIWAXS pattern of the transferred O-Cu-BHT thin film, which reveals the face-on layer orientation. (e) HR-TEM image of O-Cu-BHT thin film in a large area and the inset shows the extracted FFT pattern. (f) The ADF-STEM image of O-Cu-BHT.

Besides O-Cu-BHT film, the high uniformity and ultra-flat surfaces of the other OLGSS thin films were confirmed by the SEM images (Supplementary Fig. 5-12). Moreover, the SEM-based energy-dispersive X-ray spectroscopy (EDX) spectra and elemental mapping verified the homogeneous distribution of the elemental compositions. The Raman spectra of these samples confirmed the disappearance of signals assigned to S-H bond[39] at ~2500 cm$^{-1}$ or O-H bond[40] at ~3300 cm$^{-1}$ in the resultant $c$-MOF thin films (Supplementary Fig. 13), suggesting the efficient coordination between metals and substitution groups. As indicated by the 3D AFM topographic images (Supplementary Fig. 14), the extracted $R_q$ values of O-Ni-BHT, O-Co-BHT, O-Pd-BHT, O-V-BHT, O-Pt-BHT, O-Cu-HHB, O-Ni-HHB and O-Co-HHB thin films were determined as 3.33 Å, 3.08 Å, 2.69 Å, 3.72 Å, 3.11 Å, 4.54 Å, 4.42 Å, and 3.34 Å respectively. Taking the $R_q$ value as a guide, the surfaces of the 2D $c$-MOF thin films by OLGSS are at least 10 times smoother than those of the films by traditional LIAS and CVD methods (20 to 10$^3$ Å)[9,25–27,31–34]. The surface flatness of these OLGSS thin films can be comparable with that of the commercial dielectric SiO$_2$/Si substrates ($R_q$ = ~1.66 Å, Supplementary Fig. 15), suggesting the high compatibility with current microelectronic workflows. In addition, the optical microscopy (OM) images also validate the ultra-flat characteristic of the 9 OLGSS film samples (Supplementary Fig. 16). Moreover, the uniformity of the synthetic O-Cu-BHT thin film was also confirmed by the Raman mapping images with homogeneous intensity color (Supplementary Fig. 17). Indicated by the results of X-ray photoemission spectroscopy (XPS) measurements, no peaks assigned to Ga were detected (Supplementary Fig. 18), confirming that no chemical impurity was introduced by the OLGSS process and the subsequent transfer process.

To evaluate the layer orientation and the crystalline structure of the resultant O-Cu-BHT thin film, we performed grazing incidence wide-angle X-ray scattering (GIWAXS) analysis on a macroscopic scale. The 2D GIWAXS pattern in Fig. 2d presents an in-plane peak at $Q_{xy} = 0.85$ Å$^{-1}$ assigned to the (100) peak of O-Cu-BHT, while the intense arc at $Q_z = 1.81$ Å$^{-1}$ reveals the $\pi$-$\pi$ stacking (001) peak along the out-of-plane direction, which suggests a face-on layer orientation. Furthermore, the crystalline nature of the O-Cu-BHT thin film after transfer to the Cu grid was also probed by high-resolution transmission electron microscopy (HR-TEM, Fig. 2e), which revealed the well-preserved periodic structure across the entire ~10000 nm² area and no discernible defects in such a highly ordered 2D hexagonal lattice. The fast Fourier transform (FFT) image (inset of Fig. 2e) extracted from the HR-TEM image revealed a typical hexagonal diffraction pattern, which also revealed the face-on oriented feature of the O-Cu-BHT film. The annular dark-field scanning transmission electron microscope (ADF-STEM) image in Fig. 2f presents a lattice spacing of ~7.39 Å, which is in agreement with the (100) lattice plane of O-Cu-BHT. In addition, the crystalline structures of the other OLGSS 2D *c*-MOF films were also demonstrated by TEM imaging in Supplementary Fig. 19.

As the increase of film thickness generally engenders the amplification of surface unevenness[9,25,34], we then examine the roughness of 2D *c*-MOF thin films with varied growth duration. Taking the O-Cu-BHT as an example, the film thickness was varied from 1.75 nm to 208.91 nm with increasing the reaction time from 10 min to 12 h, (Fig. 3a and b), and the corresponding $R_q$ value over $1 \times 1$ μm² region was merely increased from 2.84 Å (Fig. 3d) to 7.13 Å (Fig. 3e). As a contrast, we also utilized traditional CVD method with solid SiO$_2$/Si wafer as substrate (termed as solid-CVD) to synthesize Cu-BHT (named as S-Cu-BHT) film with the thickness of 104.80 nm (Fig. 3c), which displayed high roughness with a $R_q$ value of 92.21 Å (Fig. 3f). The GIWAXS pattern of the S-Cu-BHT film also indicated that no preferred orientation could be observed in such an uneven film (Supplementary Fig. 20). Moreover, the Cu-BHT film synthesized by LIAS also exhibited high surface roughness with a $R_q$ value of 89.53 Å (Supplementary Fig. 21 and 22). These topographies are at least 10 times rougher than those OLGSS samples with similar thicknesses (Fig. 3g). Furthermore, we have also characterized the surface flatness of the resultant nine 2D *c*-MOF films by OLGSS with their thickness ranging from 8.81 nm to 119.79 nm (Supplementary Fig. 23-31). It is noteworthy that the measured $R_q$ values of the films remain of the same magnitude regardless of the large variations in their thicknesses, thereby validating the remarkable robustness and reproducibility of the OLGSS strategy. The thickness-dependence $R_q$ values of the 14 kinds of different 2D *c*-MOF thin films by solid-CVD, LIAS and our OLGSS methods are displayed in Fig. 3h (also see in Supplementary Fig. 32, Supplementary Tables 2 and 3).

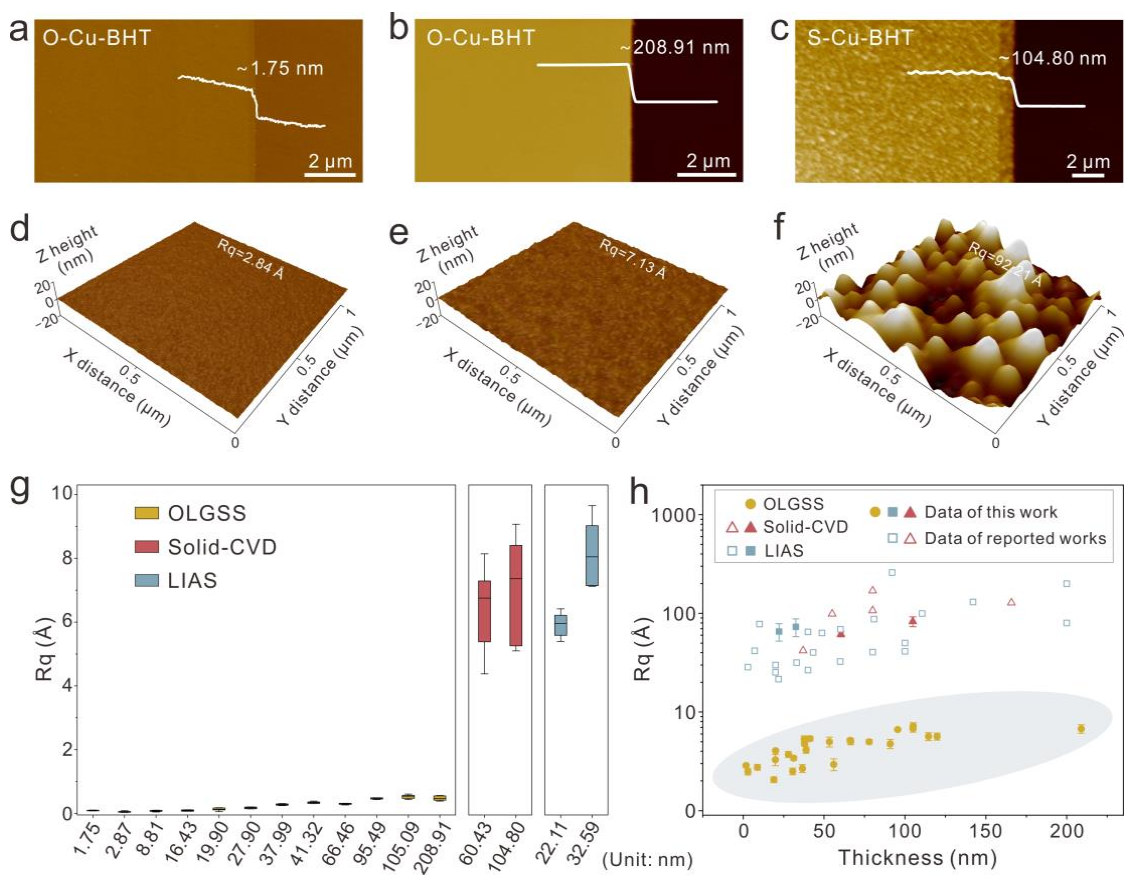

**Fig. 3 | Surface evenness of 2D *c*-MOF thin films synthesized by different methods.** (a and b) AFM images of O-Cu-BHT films with thicknesses of 1.75 nm and 208.91 nm, respectively. (c) AFM image of S-Cu-BHT film by traditional solid-CVD synthesis with thicknesses of 104.80 nm. The insets show the corresponding height profiles. (d and e) The 3D AFM topographic data of the 1.75-nm and 208.91-nm O-Cu-BHT films, respectively. (f) 3D AFM image of 104.80-nm S-Cu-BHT film. (g) Statistics of the surface roughness of the Cu-BHT thin films with film thickness synthesized from 3 different methods. The corresponding film thickness values are labeled on the x axis. (h) Comparison of the $R_q$ of 2D *c*-MOF thin films by OLGSS with those of the thus-far reported MOF films in terms of film thickness. The yellow circles, red triangles and blue squares represent the films synthesized by OLGSS, solid-CVD and LIAS methods, respectively. The solid ones and hollow ones represent the data recorded from our experiments and reported literatures, respectively (details seen in Supplementary Fig. 32, Supplementary Tables 2 and 3).

**Mechanism understanding of OLGSS.** To understand the growth mechanism, we firstly accessed the adsorption of aromatic ligands on $SiO_2$ and Ga surfaces with face-on (parallel to the substrate) and edge-on (perpendicular to the substrate) orientations by density functional theory (DFT) calculations. For instance, the DFT results reveal a much higher adsorption energy (0.773 eV/nm$^2$) of the BHT ligands with face-on orientation on the Ga surface (Fig. 4a), compared with that of the edge-on orientation (0.047 eV/nm$^2$, Fig. 4b). This result indicates that the deposition of BHT ligands on Ga surface prefers the 2D arrangement with a face-on orientation rather than an edge-on orientation, which is thereby beneficial for the subsequent formation of smooth 2D *c*-MOF thin film. In addition, we also performed the contrast calculations of the adsorption of BHT ligands on OH-terminated $SiO_2$ surface (seen in Method), which suggest that the face-on (Fig. 4c) and edge-on (Fig. 4d) orientations present comparable adsorption energies of 0.290 eV/nm$^2$ and 0.151 eV/nm$^2$, respectively. This is responsible for a disordered orientation of the BHT ligands on the $SiO_2$ substrate and the subsequent formation of rough films. To further understand the influence of the substrates on the ligand arrangement, we investigated the adsorption and assembly of BHT ligands on Ga and $SiO_2$ surfaces under CVD conditions. In a typical experimental procedure, upon heating at 130 °C without introducing metal salt, the BHT ligands could be deposited onto the substrate surface for subsequent self-assembly. The SEM images show that flat flakes were formed for the assembly of BHT ligands on the Ga substrate, while vertical rod-like crystals in random directions were observed on the $SiO_2$ surface (supplementary Fig. 33).

Next, the thermodynamic process of the 2D *c*-MOF film formation was explored for in-depth evaluation of the OLGSS mechanism. As described by the classical thin-film growth theory, the film growth behavior can be dictated by the interaction strength between the film and substrate[41,42]. Strong adhesion interaction between 2D *c*-MOF thin films and substrates can trigger the layer-by-layer growth mode, further ensuring the formation of atomically smooth films. Given this, the interface adhesion strengths of *c*-MOF films on Ga and $SiO_2$ substrates were also calculated by DFT method to probe the preferred growth modes. As indicated by the calculation results, the adhesion energy of the Cu-BHT layer on the Ga surface was calculated as high as ~3.26 eV/nm$^2$ (Fig. 4e), which was more than 2-times stronger than that of the Cu-BHT layer on $SiO_2$ surface (estimated as ~1.31 eV/nm$^2$, Fig. 4f), thus highly beneficial for the growth of ultra-smooth Cu-BHT film on Ga based on the layer-by-layer growth mode.

Antithetically, the island growth could be triggered by the inferior interface interaction, which resulted in the formation of rough Cu-BHT films on $SiO_2$. In addition, molecular dynamics simulation of the Cu-BHT@Ga at 373 K indicated that the Cu-BHT layer sustained the well-preserved 2D layered structure at the experimental temperature (supplementary Fig. 34), thus confirming the strong interaction between Cu-BHT layer and liquid Ga surface.

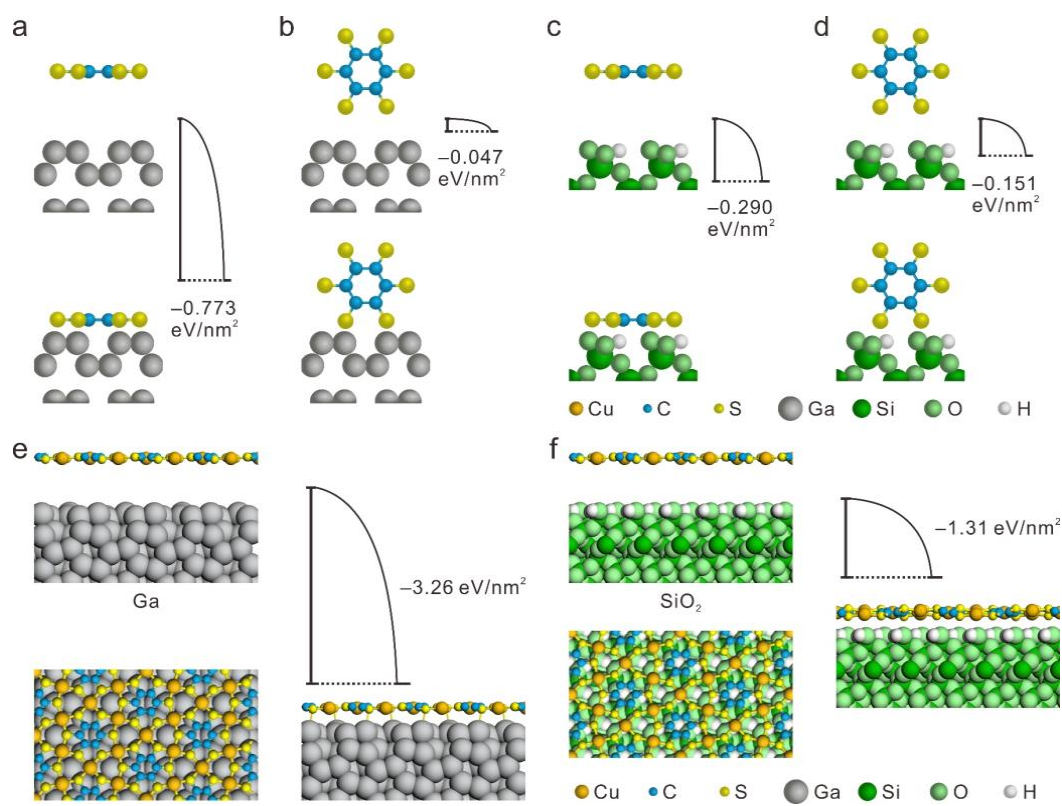

**Fig. 4 | Growth mechanism behind OLGSS strategy with O-Cu-BHT 2D *c*-MOF as a typical example.** (a and b) Simulation of the adsorption of BHT molecules with face-on and edge-on orientation on Ga surface, respectively, and the calculated adsorption energies. (c and d) Simulation of the adsorption of BHT molecules with face-on and edge-on orientation on OH-terminated $SiO_2$ surface, respectively, and the calculated adsorption energies. (e and f) Simulation of the adsorption of Cu-BHT layers on the surfaces of Ga and OH-terminated $SiO_2$, respectively, and the calculated adsorption energies.

**Device integration and electrical conductivity properties.** The ultra-smooth surfaces of OLGSS 2D *c*-MOF thin films afford us residual-free interfaces with atomic sharpness, thus allowing for the formation of ideal ohmic contacts with electrodes and low contact resistances. In this regime, lateral 2-probe devices with diverse channel lengths were fabricated via depositing Au electrodes on the ultra-smooth OLGSS 2D *c*-MOF films (Fig. 5a, supplementary Fig. 35) and the contrast solid-CVD films (grown on SiO₂/Si substrate). The contact resistances of the devices were then measured via adopting the transfer length method (TLM)[43]. Fig. 5b and c present the extracted device resistances of the O-Cu-BHT film and the contrast S-Cu-BHT film (by solid-CVD synthesis) in terms of the corresponding channel lengths, respectively, which show excellent linear fitting. The extrapolation of the fitted lines in Fig. 5d indicates that the contact resistance for O-Cu-BHT film was measured to be ~12.69 Ω (red line), which was reduced by a factor of ~13.4 compared to that of S-Cu-BHT film (~170.14 Ω, blue line). Besides, the O-Cu-BHT film exhibited an electrical conductivity as high as 1007.5 S/cm (measured by four-probe method), which is much higher than that of S-Cu-BHT film (~ 650.1 S/cm) (supplementary Fig. 36).

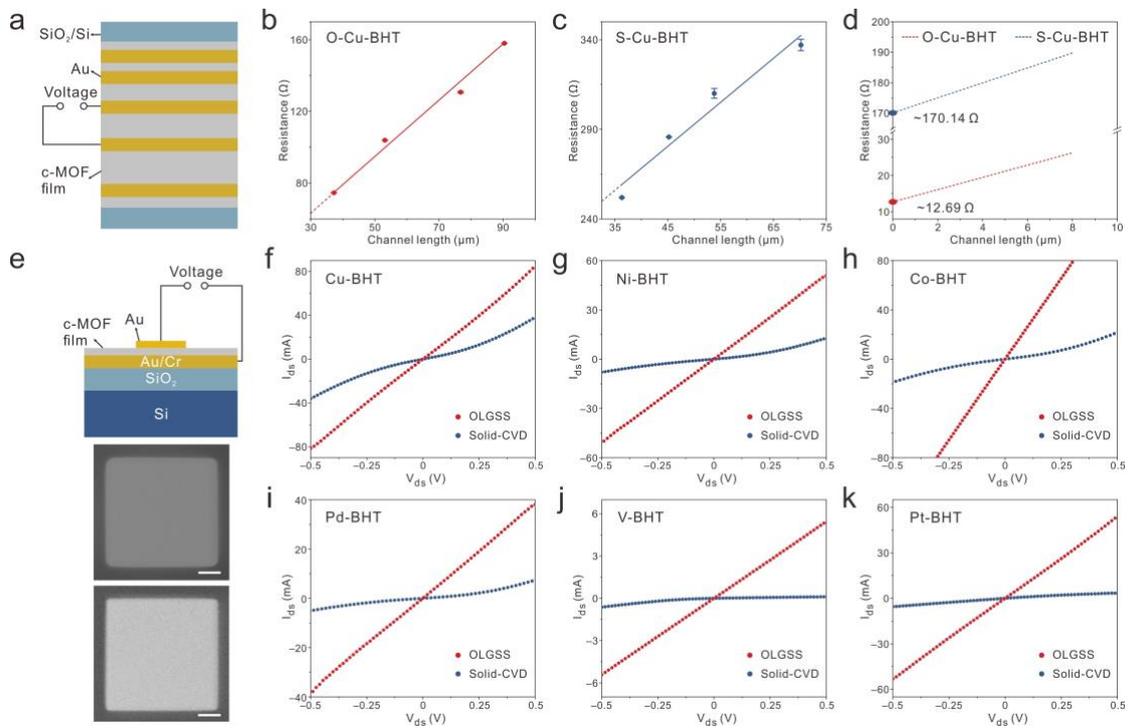

**Fig. 5 | Interfacial contact of microelectronic devices based on various 2D *c*-MOF thin films.** (a) Top-view schematic of the lateral 2-probe devices with diverse channel lengths to perform the TLM measurements. Resistance versus various channel lengths for the electrical devices fabricated by (b) O-Cu-BHT thin films and (c) S-Cu-BHT thin films. (d) Extrapolation of the fitted lines yields contact resistances of 12.96 Ω assigned to the O-Cu-BHT film, and 170.14 Ω assigned to the S-Cu-BHT film, respectively, suggesting the upgraded interface based on ultra-smooth OLGSS films. (e) Side-view schematic of the vertical device (upper panel) and the typical SEM images of vertical

devices fabricated by O-Cu-BHT film (middle panel) and S-Cu-BHT film (lower panel). The scale bars are 20 μm. (f−k) The typical I−V curves of 6 different 2D *c*-MOF thin films. Red lines represent characteristics of the OLGSS 2D *c*-MOF films, and the blue lines represent that of the solid-CVD 2D *c*-MOF films.

Furthermore, 2-probe vertical devices (the upper panel of Fig. 5e) were fabricated based on the OLGSS 2D *c*-MOF films (the middle panel of Fig. 5e) and solid-CVD films (the lower panel of Fig. 5e) to investigate the charge transport along the perpendicular direction to the substrate. Here, the 2D *c*-MOFs films with similar thickness were transferred onto the Au/Cr/SiO$_2$/Si substrate, and the Au electrodes with a size of $100 \times 100$ μm$^2$ were then deposited on the top surface of the *c*-MOF films via employing commercial TEM Cu grids as masks. Owing to the large-area contact between films and deposited Au electrodes, the I−V curves of the 2D *c*-MOF films (including Cu-BHT, Ni-BHT, Co-BHT, Pd-BHT, V-BHT, and Pt-BHT) by different synthesis methods are greatly affected by their interface quality (Fig. 5f−k). The devices based on OLGSS films present linear I−V characteristics signifying ohmic-like contact with the Au electrodes (red curves in Fig. 5f−k), whereas the nonlinear I−V curves due to the Schottky behavior were observed for solid-CVD films (blue curves in Fig. 5f−k). It is evident that the ultra-smooth surfaces of the OLGSS 2D *c*-MOF thin films translate into highly improved interface contacts of the fabricated devices, thereby significantly upgrading the compatibility with microelectronic workflows.

**2D *c*-MOF/MoS$_2$ van der Waals heterostructures.** By virtue of the ultra-smooth surface, vdWHs were constructed via wet-transferring MoS$_2$ monolayers onto the O-Cu-BHT thin film (Fig. 6a, synthesis details seen in Method). The MoS$_2$ monolayer was synthesized by CVD method (see Method) and exhibits a single-crystalline nature characterized by Raman and TEM measurements (supplementary Fig. 37 and 38). Compared to the original MoS$_2$ on SiO$_2$, apparent enhancement of PL intensity and blue-shift of PL peak were observed in the vdWHs (Fig. 6c), as suggested by the intensity mapping at 660 nm (Fig. 6b-i) and peak position mapping (Fig. 6b-ii). The statistical analysis of the PL peak positions over 100 vdWHs and MoS$_2$ samples also proves the peak shift (supplementary Fig. 39). The variation of PL spectra indicates a highly-decreased population of charged excitons (trion, originated from the *n*-type doping of CVD MoS$_2$ with S vacancies) and conversely, increased neutral excitons of the MoS$_2$ components (supplementary Fig. 40)[24,44], leading to a measured optical bandgap (~1.89 eV) approaching to that of the intrinsic $MoS_2$ (1.90 eV)[45]. The suppressed *n*-type doping should be contributed to the interlayer charge transfer[46], further revealing a tightly contacted interface between O-Cu-BHT layer and monolayer $MoS_2$ with negligible charged impurities. The weakening of Raman intensity was observed in the overlapping region (Fig. 6b-iii and -iv, supplementary Fig. 41), which was attributed to the interaction of the two components[47]. As contrast, the vdWHs comprising S-Cu-BHT layer (supplementary Fig. 42) did not exhibit noticeable alteration in PL peak position owing to the inferior interface quality.

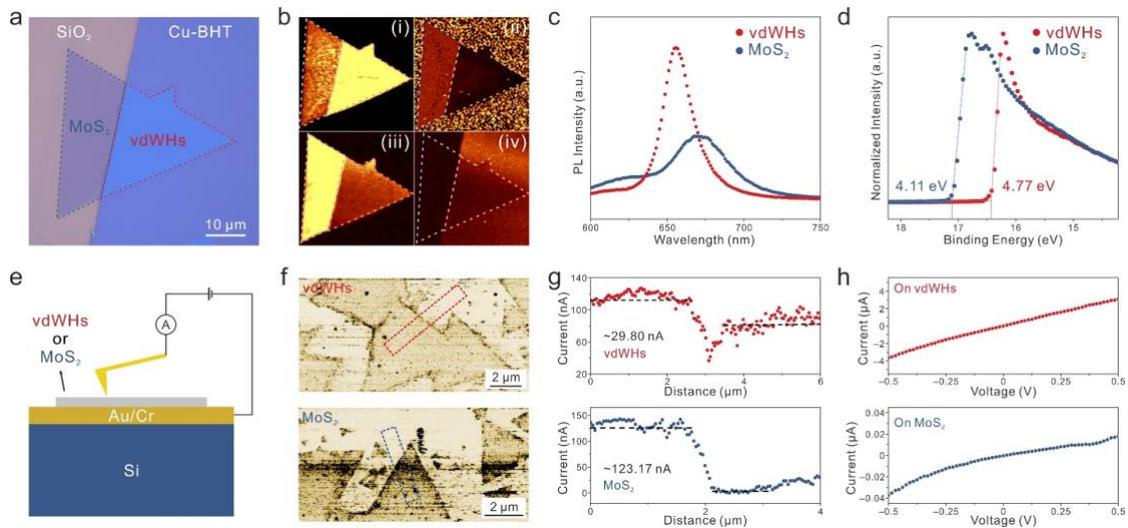

**Fig. 6 | OLGSS thin film based vdWHs.** (a) The OM image of fabricated vdWHs between O-Cu-BHT thin film and monolayer $MoS_2$ single crystal. The $MoS_2$ region is marked by a blue dashed line and the vdWH region is marked by a red dashed line. (b) The corresponding mapping images extracted from PL intensity at 660 nm (panel i), PL peak position (panel ii), $MoS_2$ Raman peak intensity at ~384 $cm^{-1}$ (panel iii) and Cu-BHT Raman peak intensity at ~1483 $cm^{-1}$ (panel iv) of the sample shown in (a). (c) Typical PL spectra extracted from the $MoS_2$ region and the vdWH region. (d) The UPS spectra of the monolayer $MoS_2$ and the vdWHs. (e) Schematic of the setup for c-AFM measurements. (f) Typical c-AFM images, (g) corresponding line profiles, and (h) extracted I–V curves of the vdWH (upper panel) and the monolayer $MoS_2$ samples (lower panel) on Au/Cr/Si substrates.

Moreover, the emerged charge transfer between O-Cu-BHT and $MoS_2$ is also expected to regulate the WF of $MoS_2$, which was probed by ultraviolet photoemission spectroscopy (UPS) (Fig. 6d) and calculated from the incident photon energy (21.22 eV) and the onset energy of the secondary electrons. The WF of the $MoS_2$ on Au/Cr/Si substrate was extracted to be ~4.10 eV and was increased to ~4.75 eV for the Cu-BHT/MoS₂ vdWHs, suggesting a large WF modulation of ~0.65 eV. Considering that the WF of clean Au is ~5.15 eV[48], the reduced Schottky barrier height led to a drastically improved electrical contacts between Au and vdWHs, as is also demonstrated by the conductive-AFM (c-AFM) measurements (Fig. 6e). The current c-AFM images (Fig. 6f) and the corresponding line profiles (Fig. 6g) clearly indicate that the vdWHs behave as a superb channel material with higher and more uniform electrical conductivity than that of single MoS₂ sample. Furthermore, the c-AFM-based I–V characteristics of vdWHs suggest a conductivity improvement of 2 orders of magnitude compared with that of MoS₂ (from ~7×10⁻⁶ Ω to ~5×10⁻⁸ Ω, as shown in Fig. 6h).

In conclusion, 9 types of ultra-smooth 2D $c$-MOF thin films with their $R_q$ values down to ~2 Å are synthesized on the liquid Ga surface by using the proposed OLGSS strategy. Such a flat surface can be reproduced in the films with their thicknesses ranging from 1.7 to 208 nm, demonstrating the high generality and reliability of the OLGSS method. DFT calculations proposed a strong interaction between 2D $c$-MOF layers and Ga, which permits the film growth followed by the layer-by-layer mode, thereby promising a uniform thin film with an enduring ultra-smooth surface. Benefiting from the ultra-smooth surface of OLGSS 2D $c$-MOF thin films, the fabricated lateral and vertical two-probe devices displayed highly elevated electrical interface contact, indicating high compatibility with current microelectronic workflows. In addition, the fabricated OLGSS 2D $c$-MOF/MoS₂ vdWHs also enables large modulations of the optical and electrical properties, including PL enhancement, PL peak shift, and work function variation. We believe that such a reliable and rational synthesis strategy will afford feasible access to versatile operable 2D $c$-MOF films with ultra-smooth surfaces as well as multicomponent MOF-based heterostructures and break one synthetic bottleneck of conductive MOFs for (opto-)electronic devices.

**Methods**

**Treatments of substrates.** The W foils (99.95%, Thermo Fisher (Kandel) GmbH) and SiO₂/Si substrates (300-nm SiO₂, MicroChemicals GmbH) were cleaned by alternately sonicating in acetone, ethanol and deionized (DI) water for 20 min. The −OH functionalized substrates were prepared by immersing the SiO₂/Si substrates in a Piranha solution, H₂SO₄:H₂O₂ = 7:3, at 80 °C for 2h. The substrates were then ultrasonicated and rinsed with DI water for $5 \times 7$ min. Before growth, all the substrates were dried in stream of $N_2$ gas.

**Synthesis of liquid Ga substrates.** A Ga droplet (~10 mg, 99.99 %, ChemPur GmbH) was placed on a W foil (1 cm × 1 cm) and heated at 1000 °C (heating rate 20~30 °C/min) for 30 min under the flow of 200 sccm $N_2$ and 200 sccm $H_2$. The heating process was performed in a quartz tube furnace (TG3, Carbolite Gero GmbH & Co. KG) and the diameter of the employed quartz tube (RAESCH Quarz (Germany) GmbH) is 26 mm.

**Synthesis of 2D c-MOF thin films on liquid Ga.** The OLGSS CVD growth of 2D c-MOFs is carried out as follows: (1) placing the precursors (put into quartz boats) and substrates at suitable positions of the quartz tube furnace; (2) purging by 200 sccm $N_2$ and then pumping the system into low-pressure (~0.5 mbar), water vapor was introduced into the system via connecting a gas washing bottle; (3) elevating the temperature of the two heating zones in 10 min; (4) dwelling in 10 min to 12 hours for achieving films with diverse thicknesses (see Fig. 2a−d); (5) cooling down to room temperature before stopping the pump. The typical setup for the OLGSS growth is illustrated in supplementary Fig. 1 and the detailed growth parameters are listed in supplementary Table 1. For the synthesis of c-MOF films by solid-CVD method, −OH functionalized $SiO_2$/Si substrates were employed as substrates for performing the growth process. The LIAS Cu-BHT thin films was synthesized at the chloroform/water interface according to the literatures[15]. Here, the BHT powders were synthesized according to the literatures[9] and the HHB powders (98%) was purchased from BLD Pharmatech GmbH. Copper(II) acetylacetonate (97%, Sigma-Aldrich Chemie GmbH), nickel(II) acetylacetonate (97%, BLD Pharmatech GmbH), cobalt(II) acetylacetonate (99%, Sigma-Aldrich Chemie GmbH), palladium(II) 2,4-pentanedionate (99.95%, 35% Pd, ChemPur GmbH), vanadium(III) 2,4-pentanedionate (97%, abcr GmbH), and platinum 2,4-pentanedionate (49.6% Pt, ChemPur GmbH) were utilized as another precursors for performing the synthesis process.

**Synthesis of $MoS_2$ single crystals.** The $MoS_2$ single crystals were firstly grown on $SiO_2$/Si substrates by a salt-assisted CVD method[49] with 100 mg S powder (99%, Grüssing GmbH) and 3 mg mixture of $MoO_3$ (99.95%, Thermo Fisher (Kandel) GmbH) and NaCl (99.5%, Fisher Scientific GmbH) ($MoO_3$:NaCl = 8:1) as precursors. The heating temperature of S powder is ~150 °C and that of Mo source and substrates is

~710 °C. After 20-min growth, the tube furnace is allowed to cooling down to room temperature.

**Transfer of OLGSS 2D c-MOFs.** The transfer process of 2D c-MOF thin films grown on liquid Ga (c-MOF/liquid-Ga) are schemed in supplementary Fig. 3 and 4. For large-area film transfer (supplementary Fig. 3), hydrochloric acid (HCl) aqueous solution (pH = 1~2) was involved. The transfer protocol consisted of followed steps: (1) gently pressing a thermal release tape (Revalpha RA-98LS(N), TELTEC GmbH) onto c-MOF/liquid-Ga system; (2) mechanically separating the tape and the large-area c-MOF films would adhere to the tape with a small amount of Ga residue; (3) floating the tape/c-MOF/residue-Ga on the diluted HCl solution for 24~48 hours to etch out Ga residue; (4) rinsing the tape/c-MOF in DI water for $5 \times 10$ min to remove the metal ions; (5) pressing the washed tape/c-MOF onto the cleaned −OH functionalized $SiO_2$/Si substrates; (6) heating the tape/c-MOF/$SiO_2$/Si at 110 °C under pressing to release tape; (7) soaking c-MOF/$SiO_2$/Si in cold acetone to remove possible residue from thermal release tape and then dry in $N_2$ stream. For avoiding acid solution, a facile transfer protocol was also proposed (supplementary Fig. 4), described as follows: (1) gently pressing a thermal release tape onto c-MOF/liquid-Ga system; (2) placing the tape/c-MOF/liquid-Ga on an ice bag to freeze Ga; (3) mechanically separating the tape and the c-MOF films would adhere to the tape without Ga residue; (4) pressing the tape/c-MOF onto the cleaned −OH functionalized $SiO_2$/Si substrates; (5) heating the tape/c-MOF/$SiO_2$/Si at 110 °C under pressing to release tape; (6) soaking c-MOF/$SiO_2$/Si in cold acetone to remove possible residue from thermal release tape and then dry in $N_2$ stream.

**Fabrication of TEM samples.** Aforementioned transfer protocols are incompatible with TEM sample fabrication as the thermal release tape will damage the supporting films of TEM grids. Thereby, the preparation of TEM sample was carried out as follows: (1) obtaining tape/c-MOF from the aforementioned two transfer protocols; (2) heating the tape/c-MOF at 110 °C (in inert atmosphere) with the c-MOF film on the top; (3) immersing released-tape/c-MOF into ethanol; (4) dispersing the c-MOF sample into ethanol via ultrasonic dispersion; (5) removing the released tape and drop-casting the dispersion onto lacey carbon TEM grids. The fabrication process was illustrated in supplementary Fig. 43.

**DFT calculations.** For the calculations of the adsorption energies between BHT molecules and two substrates, DFT as implemented in the Vienna Ab initio Simulation Package (VASP)[50,51] was used. All simulations were performed with the Perdew–Burke–Ernzerhof (PBE) exchange-correlation functional[52]. An energy cutoff of 400 eV, k-spacing of 0.5 Å$^{-1}$, and gaussian smearing of 0.3 eV were adopted. The DFT-D3 scheme[53] was used for an der Waals correction. Utilizing a conjugate gradient optimizer, the geometry was relaxed until the total (free) energy change and the band-structure-energy change was below $10^{-4}$ eV between steps. The adsorption energy ($E_{ads}$) is calculated as $E_{ads} = E_{total} - (E_{BHT} + E_{slab})$, which refer to energy of the entire structure ($E_{total}$), of the BHT molecule in vacuum ($E_{BHT}$) and of the slab ($E_{slab}$). In order to mimic the density of a liquid Ga interface, the density of the unit cell was adjusted to the one of liquid Ga at 100 °C (6000 kg/m$^3$)[54]. Then, a constant volume relaxation was performed. Based on the relaxed bulk structure, a 4 × 4 supercell Ga(001) slab was created and a vacuum layer of 20 Å is imposed. The slab is then relaxed with constant cell parameters. For the SiO$_2$ slab, the quartz structure[55] was used as initial structure and a constant volume relaxation was performed. Based on the relaxed bulk structure, a 4 × 4 supercell SiO$_2$(001) slab was created and a vacuum layer of 20 Å is imposed. The top layer was saturated with OH groups and the bottom Si atoms were saturated with H atoms. The slab is then relaxed with constant cell parameters. The interactions between Cu-BHT layers and substrates were calculated by the first-principles method with the VASP 6.3.0. PBE within the generalized gradient approximation (GGA) function was used for the treatments of exchange-correlation interaction. Spin-polarization effect was taken into account during the calculations and DFT-D3(BJ) method was employed to correct the van der Waals interactions. A $\Gamma$-centered $k$-point meshes of 3 × 2 × 1 for Cu-BHT@G$a$(001) system and 2 × 1 × 1 for Cu-BHT@SiO$_2$(001) system were adopted. The kinetic energy cutoff is 400 eV for plane-wave basis. The energy convergence criterion and force convergence criterion were set as $10^{-4}$ eV and 0.02 eV/Å, respectively.

**Fabrication of electrical devices.** For TLM and vertical devices, the Au electrodes (~50 nm) were deposited by a magnetron sputtering system (nanoPVD-S10A, Moorfield Nanotechnology Limited), where home-made masks fabricated from commercial TEM grids were employed. The Au/Cr/SiO$_2$/Si and Au/Cr/Si substrates were fabricated by thermally deposition of Cr and Au on SiO$_2$/Si and Si substrates (MicroChemicals GmbH). All the electrical measurements were performed in air by a probe station and the I−V characteristics were measured by Keysight B1500A semiconductor analyzer. No further treatment, e.g., annealing, was introduced before tests.

**Fabrication of vdWHs.** The vdWHs were fabricated via directly transferring $MoS_2$ single crystals onto O-Cu-BHT or S-Cu-BHT thin films. The fabrication process was consisted of followed steps: (1) spin-coating poly(methyl methacrylate) (PMMA, AR-P 679.04, Allresist GmbH) on the as-grown $MoS_2$ samples at 3000 rpm for 1 min; (2) releasing the PMMA/$MoS_2$ film by etching out $SiO_2$ layer in a diluted hydrofluoric acid (HF, 48–51%, Thermo Fisher (Kandel) GmbH) (HF:$H_2O$ = 1:4) and rinsing the PMMA/$MoS_2$ film in DI water for $5 \times 10$ min; (3) dissolving PMMA with hot acetone after the film was transferred onto the Cu-BHT thin films. The transfer of $MoS_2$ onto other target substrates (e.g. TEM grids) also adopted the same approach.

**Characterizations.** The AFM topographic images were recorded in the NX10 system (Park Systems) via non-contact mode at room temperature, the c-AFM was measured in the same device via contact mode at room temperature. Raman spectra and mappings were performed in a confocal Raman microscope with an excitation laser wavelength of 488 nm (Alpha300R, WITec) at room temperature. OM images were acquired from an optical microscope (Microscope Axioscope 5, ZEISS) and the Raman microscope. The four-probe electrical conductivity was measured by a SIGNATONE Pro4 system. The SEM images and corresponding EDX results were recorded by a Zeiss Gemini 500 SEM system. The TEM images were taken at an operation voltage 200kV (JEOL JEM F200 and Carl Zeiss Libra 200 Cs). To acquire HR-ADF images of the beam-sensitive MOF sample, a fast time series was acquired to reduce the dose on the sample. To compensate for drift in the image series, rigid and non-rigid registration was used by virtue of the SmartAlign software package[56]. The signal-to-noise ratio was further improved by applying template-matching algorithms to the ABS-filtered images using the SmartAlign software package for template matching. The samples were transferred to an ultrahigh vacuum chamber (ESCALAB 250Xi by Thermo Scientific, base pressure: $2 \times 10^{-10}$ mbar) for UPS and XPS measurements. UPS measurements were carried out using a He discharge lamp (hν = 21.2 eV) and a pass energy of 2 eV. XPS measurements were carred out using an XR6 monochromated Al Kα source (hν = 1486.6 eV) and a pass energy of 20 eV. The GIWAXS measurements were performed at Beamline ID10 at ESRF, Grenoble, France. The beam energy was 22 keV with a spot size of 35 µm (vertical) × 10 µm (horizontal). A Pilatus 300k area detector was positioned 418.8 mm behind the sample. The sample-detector distance and the beam center on the detector were verified by measuring lanthanum hexaboride and silver behenate as references. The incidence angle of the beam was 0.09° and the samples were exposed between 10 - 30 s to the beam depending on the intensity of the scattering signals. All data was then analyzed using WxDiff.

**Data availability**

The data that support the findings of this study are available from the corresponding author upon reasonable request.

**Code availability**

The custom code for the phase field analysis is available from the corresponding authors on reasonable request.

**Acknowledgements**

This work was financially supported by ERC starting grant (FC2DMOF, no. 852909), ERC Consolidator grant (T2DCP), SFB-1415 (no. 417590517), GRK2861 (no. 491865171), EMPIR-20FUN03-COMET, as well as the German Science Council, Center for Advancing Electronics Dresden (cfaed). This project has received funding from the European Research Council (ERC) under the European Union's Horizon 2020 research and innovation programme (ERC Grant Agreement n° 714067, ENERGYMAPS). We acknowledge the European Synchrotron Radiation Facility (ESRF) for provision of synchrotron radiation facilities and we would like to thank Dr. Oleg Konovalov NAME for assistance and support in using beamline ID10. We acknowledge Dresden Center for Nanoanalysis (DCN) at TUD. R.D. thanks the Taishan Scholars Program of Shandong Province (tsqn201909047), Natural Science Foundation of Shandong Province (ZR2023JQ005) and the National Natural Science Foundation of China (22272092). J.L. gratefully acknowledges funding from the Alexander von Humboldt Foundation.

## Author contributions

R.D. and X.F. conceived this project. J.L. and Y.C. carried out the CVD growth experiments and the Raman, PL, AFM and SEM measurements as well as the device fabrication. X.H. provided the BHT ligand. Y.R., D.B., A.D., J.G. and G.C. conducted the DFT calculations. M.D. and Y.V. performed the XPS and UPS measurements and analyzed the spectra. D.P., X.L., B.Z. and Z.L. performed the TEM measurements. M.L. contributed to the SEM measurements. M.H. and S.C.B.M. provided the GIWAX measurements and also contributed to the device fabrication. J.L., Y.C., R.D. and X.F. co-wrote the manuscript with contributions from all the authors.

## Competing interests

The authors declare no competing interests.